# HEAT CAPACITY OF OXIDE SCALE IN THE RANGE FROM 0 ºC TO 1300 ºC: GENERALIZED ESTIMATES WITH ACCOUNT FOR MOVABILITY OF PHASE TRANSITIONS

*Emmanuil Beygelzimer[1], Yan Beygelzimer[2]*
[1]*OMD-Engineering LLC, Dnipro, Ukraine*
[2]*Donetsk Institute for Physics and Engineering named after A.A. Galkin, National Academy of Sciences of Ukraine, Kyiv, Ukraine*

*Corresponding author: emmanuilomd@gmail.com, tel.: +380 (50) 368-63-42, 49000, Volodymyr Monomakh Street, 6 of. 303, Dnipro, Ukraine.

**ABSTRACT**

The known data on the heat capacity of magnetite ($Fe_3O_4$), hematite ($Fe_2O_3$) and iron (Fe) at different temperatures are approximated by formulas containing phase transition temperatures as varying parameters. This allows to take into account the effect of phase transition shifts, for example, due to impurities, lattice defects, grain sizes or high cooling rates. For this purpose, the entire target temperature range from 0 to 1300 ºC is divided by phase transition temperatures into separate intervals. The conjugation of the approximating functions between the intervals at the magnetic transition point is performed without a gap, and at the point of polymorphic transformation ($\alpha$-Fe ↔ $\gamma$-Fe) - with a finite gap of heat capacity values. For wüstite $Fe_{1-x}O$, which does not experience phase transformations, the temperature dependence of the heat capacity is approximated by a single smooth function. In combination with previously obtained formulas for the density of iron oxides and iron the proposed approximations allow us to estimate the specific mass heat capacity of oxide scale depending of its structural composition and temperature. By model calculations it is shown that at temperatures of 200 °C and 900 °C specific mass heat capacity of oxide scale practically does not depend on the percentage of its individual components and is approximately 750 and 850 J·$kg^{-1}K^{-1}$, respectively. At a temperature of about 575 °C, on the contrary, actually possible variations in the composition of oxide scale can lead to a change in its specific heat capacity from 850 to 1150 J·$kg^{-1}K^{-1}$. The obtained dependencies are recommended for use in mathematical modeling of production and processing of steel products in the presence of oxide scale on their surface.

**Keywords:** heat capacity; wüstite; magnetite; hematite; oxide scale; Curie point

## INTRODUCTION

The authors continues the description of engineering methods proposed by them to estimate in the range from 0 to 1300 °C the thermophysical characteristics of oxide scale, consisting, in general, of wüstite $Fe_{1-x}O$ ($x \approx 0.05...0.15$), magnetite $Fe_3O_4$, hematite $Fe_2O_3$, metallic iron, oxides of alloying elements and voids. The thermal expansion [1] and density [2] were considered earlier. This report presents results on the true isobaric specific mass heat capacity $c$ [J·$kg^{-1}K^{-1}$], which is defined as the value:

$$c(T) = \frac{1}{m} \cdot \frac{dQ}{dT} \qquad (1)$$

where $dQ$ [J] is the small amount of heat needed to change the temperature of a substance of mass $m$ [kg] by a small amount of degrees $dT$ [K] at constant pressure. For brevity, the parameter $c(T)$, the meaning of which is defined by expression (1), will be hereinafter referred to simply as the specific mass heat capacity at temperature $T$.

## METHODS

Known experimental and reference data on the heat capacity of the considered substances are approximated by the general methods [1], which allows to obtain mutually consistent temperature dependences of different thermophysical parameters. The basis of the methods consists in the use of reference points, through which the graph of the approximating function should pass. They include nodal points near the boundaries of the target temperature range (i.e., around 0 °C and 1300 °C) and critical points where there is a sharp change in properties. All coordinates (temperature and properties) of the reference points are set as constants, except for the temperatures of the critical points, which are included in the approximating functions as varying parameters. Accordingly, other parameters of these functions become dependent on the given values of critical temperatures. Such an approach makes it possible to coordinate the values of critical temperatures when calculating the heat capacity and other thermophysical parameters, as well as to take into account the influence of a possible shift of critical points in specific conditions. The ranges of movability of critical temperatures for above mentioned iron oxides and pure iron are summarized in [1]. Within each interval bounded by the critical temperatures, the specific mass heat capacity is described by a smooth function $c(T)$. As a reasonable simplification for engineering calculations and taking into account the works [3-5], the conjugation of the functions $c(T)$ at the point of magnetic transition is made without a gap, and at the point of polymorphous transformation - with a finite gap of heat capacity values on different sides from the critical temperature.

When the data in the primary source are presented only as enthalpy values, the authors recalculated them into values of specific mass heat capacity $c(T)$ [J·$kg^{-1}K^{-1}$] by the formula:

$$c(T) = \frac{H_{i+1} - H_i}{T_{i+1} - T_i} \qquad (2)$$

where $T_i$ and $T_{i+1}$, $H_i$ and $H_{i+1}$ are tabulated values of temperature [K] and enthalpy [J·$kg^{-1}$] from the primary source, with $T = (T_i + T_{i+1}) / 2$.

## RESULTS AND DISCUSSION

**Magnetite $Fe_3O_4$**

Empirical data on the specific mass heat capacity of magnetite (**Fig. 1**) are approximated separately in two temperature intervals: before and after the Curie point. The proposed formulas are summarized in **Table 1** and contain the Curie temperature ($T_1$) as a formal parameter. The graphs in **Fig. 2** demonstrate the response of these formulas to varying the Curie point in the possible range of 823-900 K (550-627 °C) [1]. For example, at 873 K (600 °C), the calculated value of the specific mass heat capacity of magnetite, is 947, 1024 or 1228 J·$kg^{-1}K^{-1}$ for Curie temperature 823, 848 or 900 K, respectively. In other words, for different values of the Curie point, the specific mass heat capacity of magnetite $c_{Fe3O4}$ at the specified design temperature of 873 K may differ by 281 J·$kg^{-1}K^{-1}$, i.e., by 27% relative to its value at the base value of 848 K. The results of similar evaluations for different design temperatures in the range of 150 K (±75 K from the Curie basic temperature) are shown in **Table 2**. According to these evaluations, the thermal expansion of magnetite can vary by as much as 45% due to changes in the Curie point over its real range of variation. This example demonstrates the possibility of using critical temperatures (in this case, the Curie point of magnetite) as adaptation parameters in mathematical simulation of processes in the presence of oxide scale on the surface of products.

**Fig.2** also explains the approach to selecting functions passing through independently selected nodal points with coordinates ($T_0$, $c_0$) and ($T_2$, $c_2$) and the critical point with coordinates ($T_1$, $c_1$). The type of the approximating function and values of its constants in each interval are selected based on the location of empirical points in the individual series. At the boundary between the intervals, the calculated graphs form a sharp peak in accordance with the exponential character of the heat capacity change at the magnetic transition point [3-5].

**Table 2.** Calculated effect of the Curie point of magnetite on its heat capacity

| Design temperature [K] ([°C]) | $c_{Fe3O4}$ [J·kg$^{-1}$K$^{-1}$] at different $T_1$ [K] ([°C]) | | | Range (in % of base - in italics) |
|---|---|---|---|---|
| | 823 (550) | *848 (575)* | 900 (627) | |
| 773 (500) | 1151 | *1093* | 1017 | 12% |
| 848 (575) | 1024 | *1350* | 1149 | 24% |
| 873 (600) | 947 | *1024* | 1228 | 27% |
| 900 (627) | 924 | *944* | 1350 | 45% |
| 923 (650) | 916 | *925* | 1037 | 13% |

At the basic value of Curie point for magnetite $T_1 = 848$ K (575 °C) the formulas from Table 1 are converted to a particular form ($c_{Fe3O4}$ in [J·kg$^{-1}$K$^{-1}$], $T$ in [K], the corresponding graph is shown in Fig. 1):

– in the range of 273 K $\leq T \leq$ 848 K

$$c_{Fe3O4} = -76.494 + 75.249 \cdot T^{0.4} + 310 e^{-0.016(848-T)} \quad (3)$$

– in the range of 848 K $< T \leq$ 1573 K

$$c_{Fe3O4} = 814.84 + 9.0001 \cdot 10^7 \cdot T^{-2} + 410 e^{-0.06(T-848)} \quad (4)$$

**Table 1.** Formulas for calculating the specific mass heat capacity of Fe$_3$O$_4$ ($T_1$ [K] is the Curie point of magnetite)

| Temperature interval [K] | $273 \leq T \leq T_1$ | | |
|---|---|---|---|
| Approximating function [J·kg$^{-1}$K$^{-1}$] | $c_{Fe3O4}(T) = a_0 + a_1 T^n + a_3 e^{-a_4(T_1-T)}$ | | |
| Constants | $n = 0.4$ | $a_3 = 310$ | $a_4 = 0.016$ |
| Coordinates of the reference points | $T_0 = 200$ K | $c_0 = 550$ J·kg$^{-1}$K$^{-1}$ | $c_1 = 1350$ J·kg$^{-1}$K$^{-1}$ |
| Coefficients to be calculated | $a_1 = \dfrac{c_1 - c_0 - a_3\left(1 - e^{-a_4(T_1-T_0)}\right)}{T_1^n - T_0^n}$ | | |
| | $a_0 = c_1 - a_1 T_1^n - a_3$ | | |
| Temperature interval [K] | $T_1 < T \leq 1573$ | | |
| Approximating function [J·kg$^{-1}$K$^{-1}$] | $c_{Fe3O4}(T) = b_0 + b_1 T^p + b_3 e^{-b_4(T-T_1)}$ | | |
| Constants | $p = -2$ | $b_3 = 410$ | $b_4 = 0.06$ |
| Coordinates of the reference points | $T_2 = 1600$ K | $c_2 = 850$ J·kg$^{-1}$K$^{-1}$ | |
| Coefficients to be calculated | $b_1 = \dfrac{c_1 - c_2 - b_3\left(1 - e^{-b_4(T_2-T_1)}\right)}{T_1^p - T_2^p}$ | | |
| | $b_0 = c_1 - b_1 T_1^p - b_3$ | | |

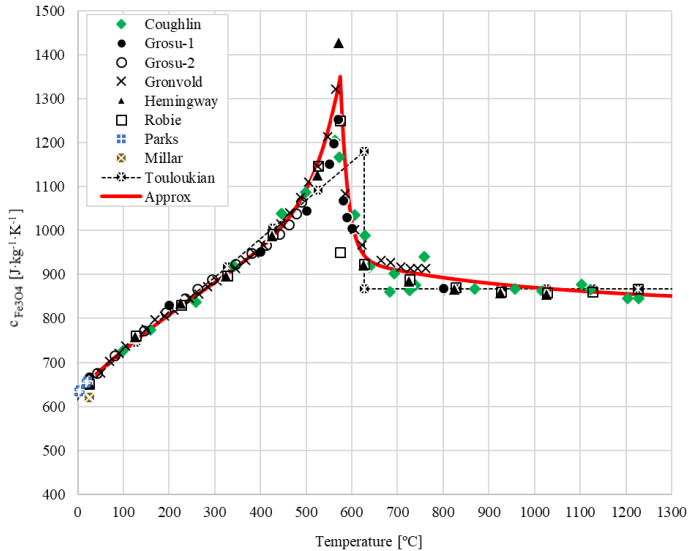

**Fig. 1.** Specific mass heat capacity of magnetite according to the empirical data of Coughlin [6] (recalculated by the authors of this article by enthalpy values), Grosu (two methods) [7], Grønvold [4], Hemingway [8], Robie [9], Parks [10, p. 114, set 1], Millar [10, p. 114, set 2] and the generalized recommendations of Touloukian [11, p. 367]. Approx - graph by formulas (3)-(4).

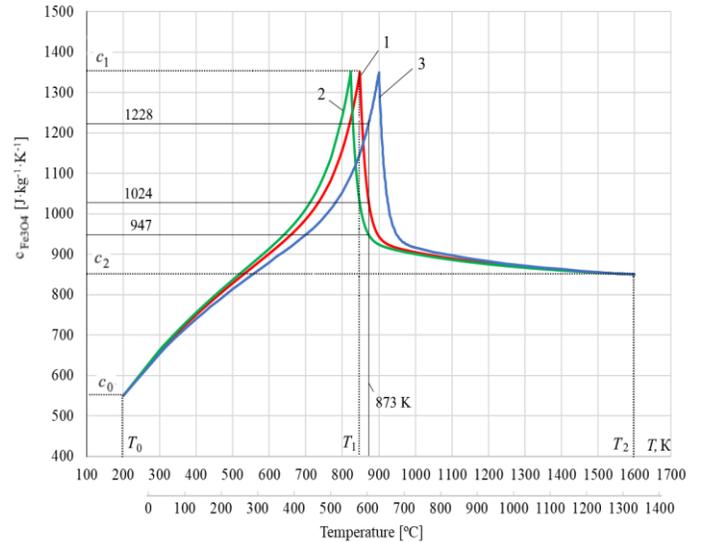

**Fig. 2.** Graphs of specific mass heat capacity of magnetite Fe$_3$O$_4$ calculated according to formulas of **Table 1** for different values of Curie point $T_1$: 1 – 848 K (575 °C); 2 – 823 K (550 °C); 3 – 900 K (627 °C).

**Wüstite $Fe_{1-x}O$**

The known data on the specific mass heat capacity of wüstite are shown in **Fig. 3**. In the entire range from 0 to 1300 °C the true specific heat capacity of wüstite increases monotonically, therefore, in contrast to approximations of thermal expansion and thermal conductivity, there is no reason to divide this temperature range by the boundary of thermodynamic stability of $Fe_{1-x}O$ (Chaudron point). The dependence of the specific mass heat capacity [J·kg⁻¹·K⁻¹] of wüstite on the temperature $T$ [K] is described by one formula in the whole range of 273 K ≤ $T$ ≤ 1573 K:

$$c_{FeO} = 548.17 + 8.7958 \cdot T^{0.5} - 556.96 \cdot T^{-2} \quad (5)$$

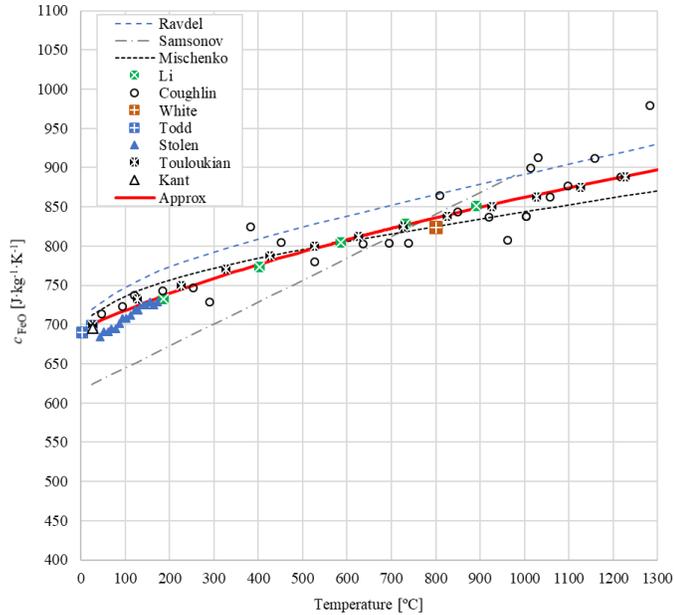

**Fig. 3**. Specific mass heat capacity of wüstite according to reference data of Ravdel [12, p. 77], Samsonov [13, p. 109], Mischenko [14, p. 14], Li [15], enthalpy measurements of White [16] and Coughlin [6] (enthalpy values were recalculated into heat capacity by the authors), experimental data of Todd [10, p. 107, set 1], Stølen [17], Kant [18] and generalized recommendations by Touloukian [11, p. 361]. The graph calculated by formula (5) is also shown (denoted as "Approx").

**Hematite $Fe_2O_3$**

When describing hematite heat capacity data (see **Fig. 4**) only one critical state, namely, magnetic transformation in the region of 950 K (677 °C), was taken into account. The second anomaly of heat capacity of $Fe_2O_3$ at 1050 K noted in [6, p. 3893] was not considered, because it was not identified in later studies [19]. **Table 3** shows the approximating formulas in general form with the Curie temperature $T_1$ as a varying parameter. The sensitivity of the specific mass heat capacity of hematite to the movability of the Curie point can be estimated from **Table 4**, which shows the values of $c_{Fe2O3}$, for some temperatures, calculated from the formulas of Table 3 at different values of the Curie point $T_1$. It can be seen that at the same design temperature, the value of the specific mass heat capacity of hematite can vary by up to 28%, depending on the position of the Curie point within its mobility range.

At the basic value for the Curie point of hematite $T_1$ = 950 K (677 °C), the formulas from Table 3 are reduced to the form ($c_{Fe2O3}$ in [J·kg⁻¹·K⁻¹], $T$ in [K], the graph is shown in Fig. 4):

– in the range of 273 K ≤ $T$ ≤ 950 K

$$c_{Fe2O3} = -31639 + 30499 \cdot T^{0.01} + 145 e^{-0.02(950-T)} \quad (6)$$

– in the range of 950 K < $T$ ≤ 1573 K

$$c_{Fe2O3} = 779.25 + 3.2687 \cdot T^{0.5} + 290 e^{-0.04(T-950)} \quad (7)$$

**Table 4.** Calculated effect of the Curie point of hematite on its heat capacity

| Design temperature [K] ([°C]) | $c_{Fe2O3}$ [J·kg⁻¹·K⁻¹] at different $T_1$ [K] ([°C]) | | | Range (in % of base - in italics) |
|---|---|---|---|---|
| | 943 (670) | *950 (677)* | 998 (725) | |
| 875 (602) | 1038 | *1031* | 996 | 4% |
| 943 (670) | 1170 | *1149* | 1055 | 10% |
| 950 (677) | 1100 | *1170* | 1065 | 9% |
| 998 (725) | 915 | *925* | 1170 | 28% |
| 1025 (752) | 895 | *898* | 980 | 9% |

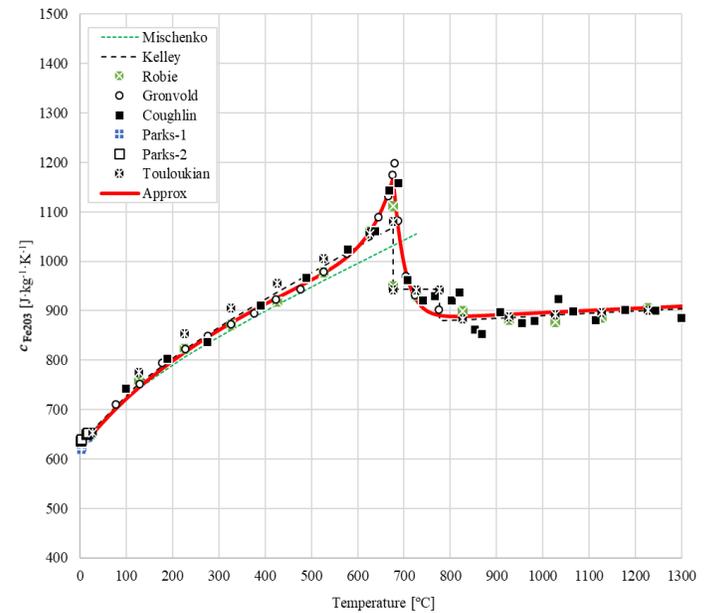

**Fig. 4.** Specific mass heat capacity of hematite according to reference data of Mishchenko [14, p. 14], Kelley [20, p. 95], Robie [9, p. 166] and the empirical data of Gronvold [19], Parks [10, p. 110, sets 1 and 5], Coughlin [6] (in the latter case the authors of this article recalculated enthalpy values into specific heat capacity by (2)) and generalized recommendations of Touloukian [11, p. 366]. For comparison, the graph calculated by formulas (6)-(7) is given (denoted as "Approx").

**Iron Fe**

Data from different authors on the specific mass heat capacity of iron $c_{Fe}$ are shown in **Fig. 5**. Approximating formulas to estimate $c_{Fe}$ [J·kg⁻¹·K⁻¹] are given in **Table 5** and contain the following parameters: $T$ [K] is the design temperature; $T_1$ [K] is the Curie point of iron; $T_2$ [K] is the temperature of polymorphic $\gamma \leftrightarrow \alpha$ transformation. A prerequisite for the application of these formulas is the ratio $T_1 < T_2$.

**Table 3:** Formulas for calculating the specific mass heat capacity of hematite Fe$_2$O$_3$ ($T_1$ [K] is the Curie point of hematite)

| Temperature interval [K] | $273 \leq T \leq T_1$ | | |
|---|---|---|---|
| Approximating function [J·kg$^{-1}$K$^{-1}$] | $c_{Fe2O3} = a_0 + a_1 T^n + a_3 e^{-a_4(T_1-T)}$ | | |
| Constants | $n = 0.01$ | $a_3 = 145$ | $a_4 = 0.02$ |
| Coordinates of the reference points | $T_0 = 200$ K | $c_0 = 520$ J·kg$^{-1}$K$^{-1}$ | $c_1 = 1170$ J·kg$^{-1}$K$^{-1}$ |
| Coefficients to be calculated | $a_1 = \dfrac{c_1 - c_0 - a_3\left(1 - e^{-a_4(T_1-T_0)}\right)}{T_1^n - T_0^n}$ | | |
| | $a_0 = c_1 - a_1 T_1^n - a_3$ | | |
| Temperature interval [K] | $T_1 < T \leq 1573$ | | |
| Approximating function [J·kg$^{-1}$K$^{-1}$] | $c_{Fe2O3} = b_0 + b_1 T^p + b_3 e^{-b_4(T-T_1)}$ | | |
| Constants | $p = 0.5$ | $b_3 = 290$ | $b_4 = 0.04$ |
| Coordinates of the reference points | $T_2 = 1600$ K | | $c_2 = 910$ J·kg$^{-1}$K$^{-1}$ |
| Coefficients to be calculated | $b_1 = \dfrac{c_1 - c_2 - b_3\left(1 - e^{-b_4(T_2-T_1)}\right)}{T_1^p - T_2^p}$ | | |
| | $b_0 = c_1 - b_1 T_1^p - b_3$ | | |

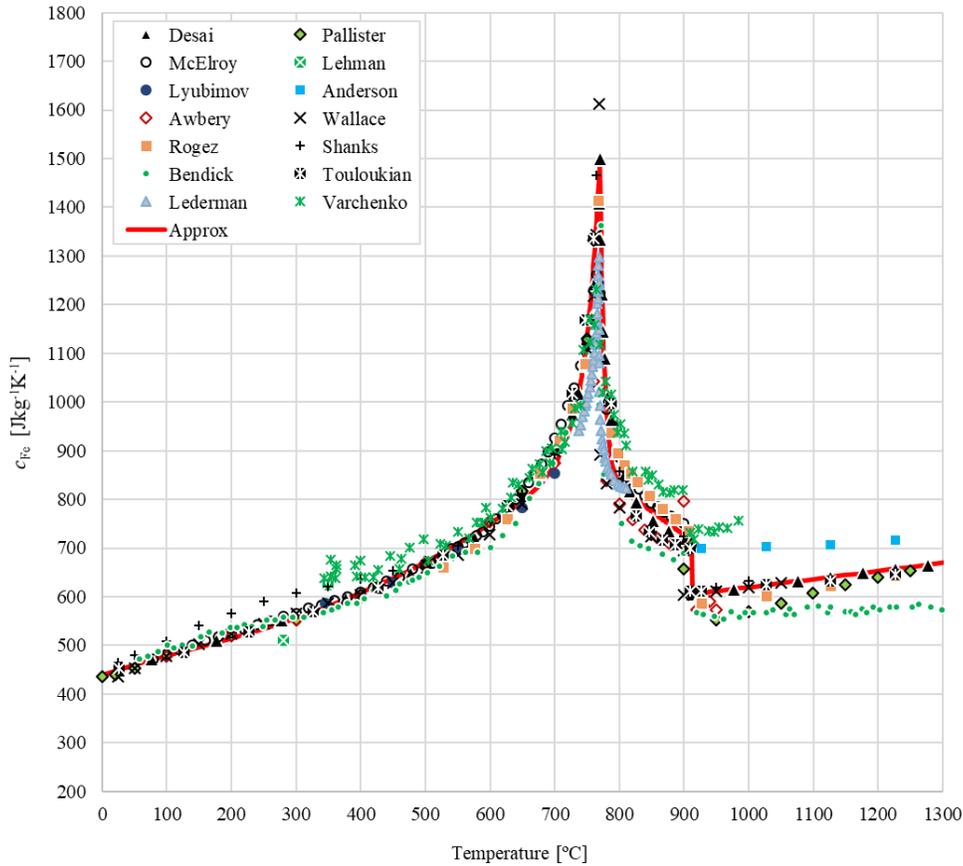

**Fig. 5.** Specific mass heat capacity of iron $c_{Fe}$ according to the experimental data of Pallister [10, p. 102, set 1], McElroy [10, p. 102, set 9], Lehman [10, p. 102, set 6], Lubimov [10, p. 102, set 5], Anderson [10, p. 102, set 12], Awbery [10, p. 102, set 4], Rogace [10, p. 102, set 6], Wallace [10, p. 102, set 8], Rogez [21], Shanks [22], Bendick [23], Ledermann [24], Varchenko [25] and generalized data from Desai [26, p. 976] and Touloukian [11, p. 292]. "Approx" is the graph of the approximating function (8)-(10).

**Table 5.** Formulas for calculating the specific mass heat capacity of iron Fe ($T_1$ [K] is the Curie point; $T_2$ [K] is the polymorphic transformation temperature)

| | | | | | |
|---|---|---|---|---|---|
| Temperature interval [K] | $273 \leq T \leq T_1$ ||||| 
| Approximating function [J·kg⁻¹K⁻¹] | $c_{Fe} = a_0 + a_1 T^n + a_2 T^m + a_3 e^{-a_4(T_1-T)}$ ||||| 
| Constants | $n = 2.7$ | $m = -2.0$ | $a_0 = 480$ | $a_3 = 580$ | $a_4 = 0.045$ |
| Coordinates of the reference points | $T_0 = 200$ K | | $c_0 = 385$ J·kg⁻¹K⁻¹ | | $c_1 = 1500$ J·kg⁻¹K⁻¹ |
| Coefficients to be calculated | $a_1 = \dfrac{c_0 - vc_1 - a_0(1-v) + a_3\left(v - e^{-a_4(T_1-T_0)}\right)}{T_1^n(u-v)}$ ||||| 
| | $a_2 = \dfrac{c_1 - a_0 - a_1 T_1^n - a_3}{T_1^m}$ ||||| 
| Auxiliary parameters | $u = (T_0/T_1)^n$ ||||| 
| | $v = (T_0/T_1)^m$ ||||| 
| Temperature interval [K] | $T_1 < T \leq T_2$ ||||| 
| Approximating function [J·kg⁻¹K⁻¹] | $c_{Fe} = b_0 + b_1 T^p + b_3 e^{-b_4(T-T_1)}$ ||||| 
| Constants | $p = 0.12$ || $b_0 = 10000$ || $b_4 = 0.15$ |
| Coordinates of the reference points | $c_{2(2)} = 716$ J·kg⁻¹K⁻¹ ||||| 
| Coefficients to be calculated | $b_1 = \dfrac{b_0 - c_{2(2)} - W(c_1 - b_0)}{WT_1^p - T_2^p}$ ||||| 
| | $b_3 = c_1 - b_0 - b_1 T_1^p$ ||||| 
| Auxiliary parameter | $W = e^{-b_4(T_2-T_1)}$ ||||| 
| Temperature interval [K] | $T_2 < T \leq 1573$ ||||| 
| Approximating function [J·kg⁻¹K⁻¹] | $c_{Fe} = d_0 + d_1(T - T_2)$ ||||| 
| Coordinates of the reference points | $T_3 = 1600$ K || $c_{2(3)} = 605$ J·kg⁻¹K⁻¹ || $c_3 = 674$ J·kg⁻¹K⁻¹ |
| Coefficients to be calculated | $d_0 = c_{2(3)}$ ||||| 
| | $d_1 = \dfrac{c_3 - c_{2(3)}}{T_3 - T_2}$ ||||| 

**Fig. 6** and **Tables 6** show the data calculated using the formulas from Table 5 to estimate the sensitivity of the specific mass heat capacity of iron $c_{Fe}$ to variations in the Curie point $T_1$ and the polymorphic transformation point $T_2$. These data show that at the same design temperature (780 ºC was chosen as an example), the values of $c_{Fe}$ can differ by 20% or more for different combinations of the Curie point and polymorphic transformation temperature within their movability range. In addition, it should be taken into account that under nonequilibrium conditions, the movability range of the polymorphic transformation temperature can be significantly greater than that taken into account in the above data.

At basic values $T_1 = 1043$ K (770 ºC) and $T_2 = 1185$ K (912 ºC), the function of the true CLTE of iron $\alpha_{Fe}$ [K⁻¹] versus temperature $T$ [K] is (graph - in Fig. 5):

– in the range of $273$ K $\leq T \leq 1043$ K:

$$\alpha_{Fe} = \left(-21 + 14.765 \cdot T^{0,14} - 7.0642 \cdot e^{-0,013(1043-T)}\right) \cdot 10^{-6} \quad (8)$$

– in the range of $1043$ K $< T \leq 1185$ K:

$$\alpha_{Fe} = \left(16.004 - 5.0041 \cdot e^{-0.05(T-1043)}\right) \cdot 10^{-6} \quad (9)$$

– in the range of $1185$ K $< T \leq 1573$ K:

$$\alpha_{Fe} = 23,0 \cdot 10^{-6} \quad (10)$$

**Table 6.** Calculated values of specific mass heat capacity of iron $c_{Fe}$ [J·kg⁻¹K⁻¹] at the design temperature $T = 1053$ K (780 ºC) for different combinations of the Curie point $T_1$ and polymorphic transition temperature $T_2$

| $T_2$ [K] ([ºC]) | $T_1$ [K] ([ºC]) | | |
|---|---|---|---|
| | 1032 (759) | 1043 (770) | 1046 (773) |
| 1183 (910) | 872 | 989 | 1072 |
| 1185 (912) | 874 | 990 | 1073 |
| 1208 (935) | 894 | 1006 | 1086 |

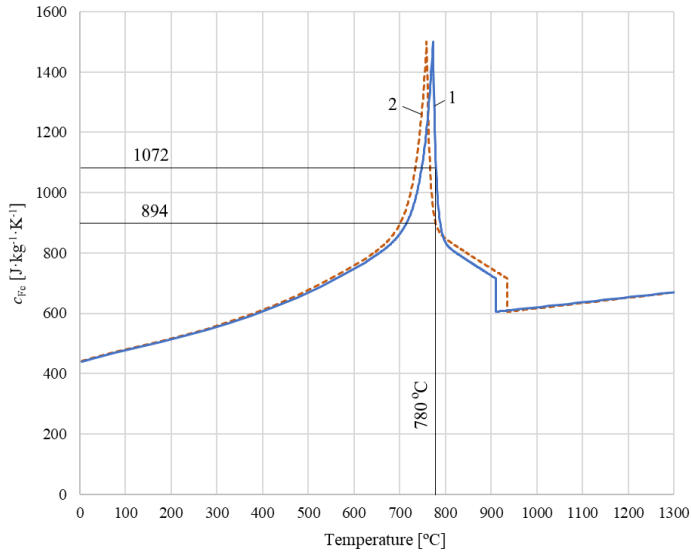

**Fig. 6.** Graphs for specific mass heat capacity of iron calculated by the formulas from Table 6 with two different combinations of the Curie point $T_1$ and the polymorphic transformation temperature $T_2$: 1 - $T_1 = 1046$ K (773 °C) and $T_2 = 1183$ K (910 °C); 2 - $T_1 = 1032$ K (759 °C) and $T_2 = 1208$ K (935 °C).

**Oxide scale as a whole**

**Fig. 7** compares the temperature dependences of the specific mass heat capacity for different structural components of the scale, calculated using the above formulas for the basic values of critical temperatures.

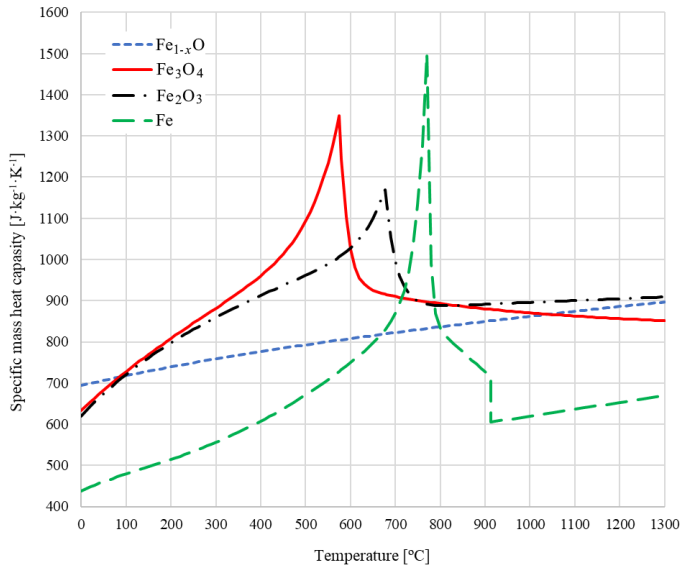

**Fig. 7.** Comparison of calculated graphs of specific mass heat capacity of wustite ($Fe_{1-x}O$), magnetite ($Fe_3O_4$), hematite ($Fe_2O_3$) and iron (Fe) as a function of temperature. The calculations were performed according to formulas (5), (3)-(4), (6)-(7) and (8)-(10) respectively

The specific mass heat capacity of scale as a multicomponent material can be defined as:

$$c_{sc} = \frac{\Delta H}{m_{sc}} = \frac{m_1 c_1 + m_2 c_2 + \cdots}{m_{sc}} = \varepsilon_1 c_1 + \varepsilon_2 c_2 + \cdots \quad (11)$$

where $m_{sc}$ - mass of a given amount of oxide scale, $\Delta H$ - change in heat content of this mass when its temperature changes by one degree, $m_1, m_2, \ldots$ - masses of individual components of oxide scale, $c_1, c_2, \ldots$ - specific mass heat capacity of the components; $\varepsilon_1, \varepsilon_2, \ldots$ - mass fraction of the components.

The mass fractions of the components $\varepsilon_i$ are related to their volume fractions $\psi_i$ by the following relations ($i = 1, 2, \ldots$):

$$\varepsilon_i = \zeta_i \psi_i \quad (12)$$

where the volume fractions are understood without taking into account the pores, i.e.

$$\psi_1 + \psi_2 + \cdots = 1 \quad (13)$$

and the symbol $\zeta_i$ denotes the relative density of this component as the ratio of its density $\rho_i$ to the true (without considering the pores) density of the oxide scale $\rho_{sc}$:

$$\zeta_i = \frac{\rho_i}{\rho_{sc}} \quad (14)$$

Taking into account (12) and the adopted notations of the scale components, expression (11) can be written in the form:

$$c_{sc} = (c\zeta\psi)_{FeO} + (c\zeta\psi)_{Fe3O4} + (c\zeta\psi)_{Fe2O3} + (c\zeta\psi)_{Fe} + (c\zeta\psi)_{xO} \quad (15)$$

where each summand is the product of the specific mass heat capacity of a given component by its relative density and volume fraction in the scale (the index $xO$ stands for the oxide of the alloying element).

The specific mass heat capacity of iron oxides and metallic iron can be calculated by the formulas described above, and the density by the formulas given in [2].

As an example, **Fig. 8** shows graphs $c_{sc}$, which are computed by the formula (15) for four different compositions of oxide scale listed in **Tables 7-8**. The first three cases conventionally assume the same component content over the entire temperature range (which can be considered in the first approximation as corresponding to the conditions of rapid cooling), and the fourth case has a variable content depending on temperature (modeling the conditions of slow cooling with decomposition of wüstite into an eutectoid mixture of magnetite and metallic iron). In all cases of calculation, the oxides of alloying elements in the composition of the oxide scale are not taken into account.

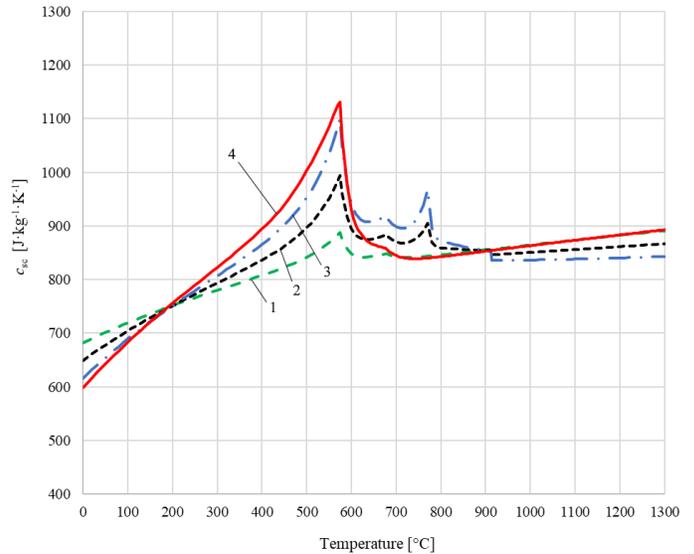

**Fig. 8.** Graphs of temperature dependence of specific mass heat capacity of scale calculated by formula (15) through the corresponding properties of its structural components. Numbers in the curves are numbers of calculated compositions of scale according to Tables 4-5.

Performed results show that there are two temperature regions in which the value of the specific mass heat capacity of oxide scale practically does not depend on its structural composition (in the real range of its change): $c_{sc} \approx 750$ J·kg$^{-1}$·K$^{-1}$ at 200 °C, and $c_{sc} \approx 850$ J·kg$^{-1}$·K$^{-1}$ at 900 °C. On the other hand, at temperatures around 575 °C (the basic value of the Curie temperature of magnetite), the specific mass heat capacity of oxide scale is unstable and can vary from 850 J·kg$^{-1}$·K$^{-1}$ to 1150 J·kg$^{-1}$·K$^{-1}$, depending on the percentage of its structural components.

In the end, we note that the porosity of the scale has no effect on its specific mass heat capacity (this is obvious from formula (11), if we consider the pores as a component with zero mass).

**Table 7.** Constant compositions of scale No. 1-3, adopted in the calculations of the graphs in Fig. 10.

| Composition | Component volume fraction $\psi$ | | | |
|---|---|---|---|---|
| | FeO | $Fe_3O_4$ | $Fe_2O_3$ | Fe |
| 1 | 0,8 | 0,15 | 0,05 | 0 |
| 2 | 0,5 | 0,35 | 0,1 | 0,05 |
| 3 | 0,2 | 0,55 | 0,15 | 0,1 |

**Table 8.** Variable composition of scale No. 4, adopted in the calculations of the graph in Fig. 10 (more details - in [1])

| Temperature, °C | Component volume fraction $\psi$ | | | |
|---|---|---|---|---|
| | FeO | $Fe_3O_4$ | $Fe_2O_3$ | Fe |
| 1300 | 0,880 | 0,100 | 0,020 | 0 |
| 1000 | 0,880 | 0,100 | 0,020 | 0 |
| 900 | 0,880 | 0,100 | 0,020 | 0 |
| 800 | 0,875 | 0,100 | 0,025 | 0 |
| 700 | 0,779 | 0,166 | 0,055 | 0 |
| 600 | 0,416 | 0,489 | 0,095 | 0 |
| 570 | 0,286 | 0,596 | 0,118 | 0 |
| 500 | 0,068 | 0,749 | 0,090 | 0,093 |
| 400 | 0 | 0,788 | 0,090 | 0,122 |
| 300 | 0 | 0,788 | 0,090 | 0,122 |
| 100 | 0 | 0,788 | 0,090 | 0,122 |

**CONCLUSION**

Known from technical literature data on the specific mass heat capacity of magnetite $Fe_3O_4$, hematite $Fe_2O_3$ and pure iron Fe are approximated by analytical functions containing the temperatures of magnetic and polymorphic transformations as varying parameters. Accordingly, the entire target temperature range from 0 to 1300 °C for each of the above components of oxide scale is divided into separate intervals with moving boundaries. In each of these intervals, the specific mass heat capacity was described by a smooth function that takes into account the exponential growth of this parameter to the Curie point. The conjugation of functions for adjacent intervals at the Curie point was performed without a break in the heat capacity, and at the point of polymorphic transformation (for Fe) with a finite break in the heat capacity. For the wüstite $Fe_{1-x}O$, which does not undergo phase transformations throughout the indicated temperature range, the generalized temperature dependence of the specific mass heat capacity was described by a single smooth function.

The data obtained allow us to calculate the specific mass heat capacity of scale $c_{sc}$ as a multicomponent mixture. The results of model calculations presented in the paper show that at temperatures of 200 °C and 900 °C the oxide scale specific heat capacity practically does not depend on the structural components content and is approximately 750 J·kg$^{-1}$·K$^{-1}$ and 850 J·kg$^{-1}$·K$^{-1}$ respectively. In the region of 575 °C, on the contrary, it depends strongly on its percentage composition, varying from 850 J·kg$^{-1}$·K$^{-1}$ to 1150 J·kg$^{-1}$·K$^{-1}$.

The proposed formulas are recommended for use in mathematical simulation of production and processing of steel products in the presence of surface scale.